# Towards A Time Based Video Search Engine for Al Quran Interpretation

Maged M. Eljazzar[1, a)], Afnan Hassan[2, b)] and Amira A. Al-Sharkawy[3, c)]

[1, 2, 3]Faculty of Engineering, Cairo University, Cairo, Egypt

a) mmjazzar@ieee.org
b) afnan_saleh@feng.bu.edu.eg
c) amira.alsharkawy@eri.sci.eg

**Abstract**. The number of Internet Muslim-users is remarkably increasing from all over the world countries. There are a lot of structured, and well-documented text resources for the Quran interpretation, Tafsir, over the Internet with several languages. Nevertheless, when searching for the meaning of specific words, many users prefer watching short videos rather than reading a script or a book. This paper introduces the solution for the challenge of partitioning the common Tafsir videos into short videos according to the search query and sharing these result videos on the social networks. Furthermore, we provide the ability of user commenting on a specific time-based frame on the video or a specific verse in a particular subject. It would be very valuable to apply the current technologies on Holy Quran and Tafsir to easy the query for verses, understanding of its meaning, and sharing it on the different social media.

## 1. NTRODUCTION

With the increase in use of social networks and mobile application, users are currently able to browse a great number of useful resources in a short time. It became easier to search for new information from the web rather than reading a book. Besides you can ask your question in several websites such as Quora and you will find an answer from experts. Many NGOs and companies make use of the huge advance on web technologies and they developed revolutionary educational platforms for sharing free learning resources. At the current time, it is obvious that millions of students study online courses in Massive Open Online Courses (MOOC) websites such as Coursera, EDX, and etc.

Several reasons encouraged students prefer studying in MOOCs rather than reading books. Firstly, each topic is consist of several sub-topics and each sub-topic is explained in a short video. Secondly, linking questions to these videos as comments or independent discussions forums. Using these boards, the user can check another question related to that topic. The last reason, several translations was proposed to the sub-video by the students. In short, the student can check a specific topic and check the asked questions in two to three minutes. Which in turn made the student very satisfied.

In terms of Islamic application on the web, Multiple Islamic applications were developed for specific needs. Most of them are related to the Quran or Hadith. Some of these applications based on the cloud as a service, while others developed for offline users. Quran – KSU Electronic Moshaf Portal Project [1] is one of the best comprehensive online portal on Quran. As users can view scanned (soft) copy of real printed Quran. Besides it produces an online recitation by many famous reciters and an advanced search feature. Another Quran Memorization program [2] was developed for memorization with good analytics to track the performance. Currently there is an increasing number of Islamic application on play store or even on the cloud [3-8].

According to internet live stats [9] around 40% of the world population has an internet connection today. Many challenges and needs arose with this huge percentage in the age of information. During Prophet Mohamad age and after it Islamic resources used to be spread by listening to a trusted teacher who learned from another teacher chained to Prophet Mohamad. Islamic Video lectures now gained great importance for those whom learning Islam even if they are Arabic native or not as it is easier to understand and simpler, but searching for certain information in this very hard.

This work introduces an approach to search for certain piece of information in Al-Quran Interpretation Videos in a few second. Besides linking short videos to questions related to the same Verses and more illustration for the same Verses on another Islamic resources such as Heritage books. At last, storing users' quarries to add more information about the trends to detect users' needs to keep enhancing the model and social awareness.

## 2. LITERATURE REVIEW

For Muslims, Quran is the revelation from God sent to mankind through Prophet Muhammad (peace be upon him). Reading Quran Interpretation is a necessity to understand the Quran meanings. According to that, Scientists spent huge time for Quran Interpretation. Currently, anyone has an internet access can easily find Interpretation for any Verses on YouTube. Besides, there are several websites introduce libraries for Quran Interpretation to different scientists with several languages such as Islamway, islamicity, http://www.altafsir.com/. Some of these websites produce the Quran Interpretation according to heritage books or even as video available to download.

A deep analysis was made in [10] to introduce a Cloud-based Portal for the Quran not only for Interpretation, but it includes Memorization, Reading with Tajweed (rules of recitation of the Quran), Recitation and more. A data mining review to compare Quran web portals in [11] to obtain awareness of Islam. Other application provides Quran Interpretation as a desktop application such as Shamla library [12] which considered one of the most resourceful Islamic libraries. Shamla contains hundreds of Quran Interpretation, which introduce the Interpretation in several books. In terms of Android users, the number of application is increasing with several languages. In [13] a good Analysis of Android based Holy Quran Applications were made listing the common applications.

It is obvious the huge contributions from different developers, or scientists in enhancing searching algorithms, data mining, digitizing the Quran Interpretation books, or developing web portals to enhance usability or user interface. This work tried to make a contribution in Searching for certain piece of information in Al-Quran Interpretation Videos. Besides, producing more explanation using the common Quran Interpretation on the other website.

## 3. METHADOLOGY

Traditional video retrieval based on visual feature extraction cannot be easily applied to Quran Interpretation videos due to the homogeneous scene composition of lecture videos. This work aimed to provide a simple user interface and well known Quran Interpretation playlist. Many challenges appeared in terms of the video quality, full playlist, and easy pattern to detect. At least we used Shaikh El Sharawy [14] library as it is well organized, good quality, and the Shaikh in the beginning of every video mention the verse and some key words for the video. Several methods were discussed to organize search query as follow:

1. Dividing the Al-Quran Interpretation Video by Al-Quran Verses and search by verse number.
2. Automated searching for certain speech pattern in the Video which will be more applicable and generalized for different videos than the first one.

The main issues when using the first method, firstly several videos on YouTube could be for the same verse number for the same scientist. Secondly, some verse number Interpretation are too long. At last, the user may want to search for a specific meaning. He is not restricted to a specific verse number.

In The Application Programming Interface (API) world, many libraries have a variety of libraries for speech recognition. In [15] a good comparison between Google, Nuance, AT&T, WIT, and IBM Watson. He used the same audio file which contains more than 3000 different phrase with considering environmental conditions like noisy environment. Then he compares APIs with the number of the extracted phrases the more the better, and wards need to be edit fewer words need to be edit is better. He measured the performance by minimum number of single character edit. Then he found that that google API is the best as it achieved 73.3% of detected phrases and 15.8% phrase need to be edit. Nuance came in second place by a large margin. In Nuance, 44.1% of the phrases were recognized perfectly and the word error rate was 39.7%. IBM (46.9.3% and 42.3% word error rate) came in third place. At last AT&T and WIT had the exact same word error rate - 63.3%, with a small advantage in exact recognition by AT&T (32.8% vs 29.5%, WIT).

Gaida et al. [16] provided a large-scale evaluation for three open-source toolkits of speech recognition: Their experiments proved that the most advanced toolkit is Kaldi [17]. However Kaldi have the shortest time and the best results, its computational cost was the highest among the others. The runner-up was CMU Sphinx toolkit [18], [19] that provided good results in a short time.

Adorf [20] provided a review from different perspectives about: Web speech API, how it works, and how web application benefit from it. Furthermore, the research provided an evaluation about the performance of Google speech API. Web speech API is an experimental javascript API that implemented by W3C API community for speech recognition and synthesis; it facilitate web application to translate speech to text.

In this work the sound pattern was detected automatically for the previously mentioned issues in the first method. Besides using speech recognition API simplify the process and performs good results. Although dividing the videos manually will perform robust results, but more time is needed.

# 4. MODEL

The model based on two main components. The first is converting the Interpretation speech into text. The second is providing a well-designed user interface. In terms of speech recognition, there are many techniques for such as Hidden Markov Model, speech recognition API, Support Vector Machine, Artificial neural network and Deep neural network, but google Arabic speech recognition was the best model for several reasons. Firstly, the application interface is simple. Secondly the Strong support for the Arabic language more than the other models. At last the speed and quality is far better than the other models. In [21] the steps of the system described in details. We followed the same steps with few modifications.

1. Collecting data from different environments.
2. Applying Acoustic Modeling.
3. Training data became text queries.
4. Building lexicon model.

The steps of the system will be collecting the dataset then segment it and training it to generate Acoustic model. One of the challenges is the number of Arabic datasets is very limited. According to that building new dataset was a necessary step. The dataset are related to the same topic to guarantee the results would be accurate in high percentage. After that, we will be able to generate the lexical model and both of Acoustic model and lexical model will generate speech encoder which will input Acoustic features and output will be hypothesis.

In this work, the previous steps were followed in sequence. In addition, linking video into text stamped by time and search for the required text in this stamped text file. Besides enable the user feedback as shown in figure 1. Every user leave a feedback for the best time based in the video. This could help the database to be more accurate time based video. In next queries search the data base consider the node (speech recognition transformed into text) and the user feedback to perform the query. Moreover storing user comments on specific time based videos per second could be useful for other users. For example if a specific video related to another video the user could provide a link on the time frame that mentioned in the video.in addition, some verse tells us a specific scientific discovery. It would be very useful sharing what science discovered after Holy Quran more than 1400 years.

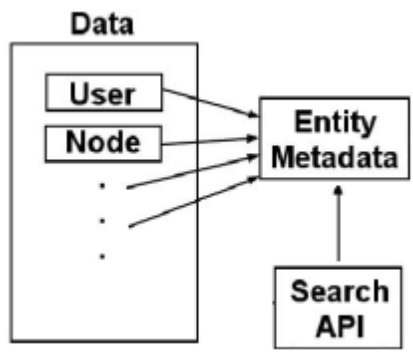

**Figure 1.** How the user can enhance search results.

This research is using Heroku Cloud Service to host the database and its application. The application is using SQLite database as a prototype. This research uses mainly JavaScript for and JSON for communication between the user and the remote database in Heroku. The user interface include only a search panel ask the user to search for a specific meaning in Quran. After the user perform a search transaction.

The remote database store the search results for further analysis. Then the database will back to the user as shown in figure 2. Firstly, the database select the video from the database. Secondly the database checks if there are any comments of other users related to time frame. At last search for the same meaning on other website such as islamweb and provide a summary or recommendation based on the search results.

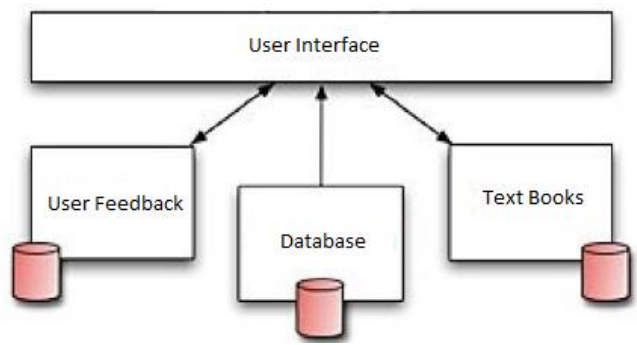

**Figure 2.** User Interface.

## 5. EXPERIMENTAL RESULT

This research aims making the search of Quran Interpretation in short time. There are huge resources for Quran Interpretation, but it lacks indexing. The test results shows a good practice for the engine. The engine results were accurate based on the video based and other connected resources stated above. Three main issues appeared during the testing:-

1. Dividing videos may miss some words.
2. Google speech recognition API need high quality audio.
3. Google speech recognition API distinguish the requests better than the speech

Firstly issued was solved by cutting overlapped Audio to get missing words. Secondly the video quality, Google search API needs high quality audio to recognize. The available Arabic data on the other hand are sampled at 8kHz. These genres and acoustic conditions do not meet the requirements for getting the best performance during the training model. It is better if the data 16kHz sampled to avoid mismatching. In terms of Google's requests. In this system, users speak their search queries, typically using a mobile phone, and the system returns a transcription and web search results. The system is trained over millions of transactions which in turn represented in the performance. Table 1 summarize the current progress, the raise in word error rate due to the quality of audio. It is clear that, when training the model using for the same Shekih, it enhance the results. At last, as the system respond to the query based on the speech recognition database and feedback system. It is expected the word error rate will decreased on previous called queries on the database.

TABLE 1. Current progress of Acoustic Model Training

| Training | Word error rate |
|---|---|
| Google speech API | 22.5% |
| IBM Watson API | 17.7% |

## 6. CONCLUSION AND FUTURE WORK

This research introduces a search engine to help internet users to reach a specific words of Quran in short time based on time of speech according to the Interpreter.

One of the objective of the system is to analysis the user search queries and its results for future analysis. This could be useful for Islamic communities in terms of the increasing awareness in the trend of issues of Islamic societies. In addition, linking the user comments to a specific time frames. This could be helpful providing comparison to other videos or more scientific facts discovered by Quran. The future work will include testing more videos to increase the number of available videos in the database. In addition, enhancing the current user interface and adding more testing to the system. In terms of the model training, it will include general speech not only requests to enhance the system performance.

At last, the future work will include considering other language than Arabic. The expansion of this research work will include the development of an Application Program Interface (API) for the database to assist developers in Islamic communities to use the database.

## ACKNOWLEDGEMENTS

The official website for Shekih Sharawy; the website introduce the whole Quran Interpretation in a simple user interface and easy to use access.

## REFRENCES